\documentclass[a4paper, 12pt]{article}

\usepackage{amsmath}
\numberwithin{equation}{section}
\usepackage{graphicx}
\usepackage{url}
\usepackage{amsbsy,amssymb} 
\usepackage{amsmath}
\usepackage{abstract}
\usepackage{hyperref}
\usepackage{color}
\usepackage{authblk}

\usepackage{fancyvrb}

\newcommand{\codeMark}[1]{{\protect\Verb+#1+}} 
\newcommand{\commandMark}[1]{{\protect\Verb+#1+}} 
\newcommand{\classMark}[1]{{\protect\Verb+#1+}} 
\newcommand{\configOptionMark}[1]{\textit{#1}}
\newcommand{\dirFileMark}[1]{{\it #1}}

\newcommand{\GeV}{\textrm{GeV}}
\newcommand{\mikrob}{~\mathrm{\mu b}}
\newcommand{\milib}{~\mathrm{mb}}
\newcommand{\dif}{\mathrm{d}}

\title{GenEx: A simple generator structure for exclusive processes in high energy collisions}
\author[1]{R.A.Kycia\thanks{e-mail: rkycia@pk.edu.pl}}
\author[2]{J.Chwastowski}
\author[2]{R.Staszewski}
\author[2]{J.Turnau}
\affil[1]{Tadeusz Ko\'sciuszko Cracow University of Technology}
\affil[2]{Henryk Niewodnicza\'nski Institute of Nuclear Physics, PAN, Krak\'ow}

\date{}
\begin{document}
\maketitle
\begin{abstract}
\noindent
A simple C++ class structure for construction of a Monte Carlo event generators which can produce unweighted events within relativistic phase space is presented. The generator is self-adapting to the provided matrix element and acceptance cuts. The program is designed specially for exclusive processes and includes, as an example of such an application, implementation of the model for exclusive production of meson pairs $pp \rightarrow p M^+M^- p $ in high energy proton-proton collisions.
\end{abstract}

\section{Introduction}
Commonly used method of testing hypotheses concerning the structure of matrix elements for inelastic processes in high energy physics is Monte Carlo simulation. One generates random ``events'' within the kinematic phase space (conserved energy and momentum) and associates with each event a weight which is a product of the square of modulus of the matrix element and a kinematic factor \cite{Generators_General}, \cite{ThePEG}. These properly weighted MC events may be processed in the same manner as real events and may yield theoretical distributions directly comparable to experimental ones. This method becomes inefficient when for dynamical reasons a non-negligible weight is observed only in the very limited regions of the phase space. For example, in high energy multi-particle production the transverse momenta of final state particles are small while their longitudinal momenta increase rapidly with collision energy. Methods for generation of events within the so-called longnitudinal phase space were developed 
long time ago \cite{Pene},\cite{Kittel},\cite{Jadach_cylindrical} and are standard components of any event generator for high energy physics. However, there is growing interest in investigation of the high energy exclusive processes (for recent review see \cite{Harland-Lang}) which are governed by sharply peaked matrix elements and nonzero weight events occupy an extremely small spot in the available phase space. There exists several event generators specially designed to this type of processes. GenEx, the event generator described in this paper, achieves a high efficiency due to  combination of a proper choice of the integration variables with self-adapting Monte-Carlo \cite{Jadach-FOAM} implemented in TFoam class of ROOT \cite{ROOT_site}. The usage of this self-adapting Monte-Carlo technique and modular C++ structure specially designed to facilitate addition of new phase space generation methods and new processes (matrix elements) are main features which make it in a class of functionality different from 
other generators 
for exclusive processes: DIME \cite{Dime}, FPMC \cite{FPMC}, ExHuME \cite{ExHuME}. In addition, GenEx, unlike ExHuME and FPMC, does not use the parton structure of the colliding particles from the beginning, what makes it suitable for modelling soft, Regge-like processes. The program extensively uses existing commonly available open source ROOT library developed at CERN \cite{ROOT_site}, in particular Foam \cite{Jadach-FOAM} and some other free software. For other general purpose software that uses Foam see, e.g., \cite{Slawinska-general}.
\par
The following sections contain general description of the generation method, program and class structure and detailed instruction how to install and use GenEx, including the examples and standard tests. In the Appendix some details of the phase space calculation are explained.

\section{General description of the event generator }\label{general}

A typical path of the Monte Carlo event generation has the following steps:
\begin{itemize}
\item create vector of random numbers $\mathcal{R}$,
\item transform $\mathcal{R}$ into a set of final state particles four-momenta $P$ which fulfil the energy-momentum conservation,
\item evaluate the matrix element square and the event total weight,
\item use the rejection method to generate events with unit weight.
\end{itemize}
Basing on general properties of a certain class of matrix elements, e.g., for multiple particle production in high energy collisions, importance sampling methods are usually employed to increase the generator efficiency. The importance sampling is performed either at the level of the first step (non-uniform random number distribution) or the second step (choice of variables which are uniformly generated) or both as in \cite{Jadach_cylindrical}. The generator described in this paper includes one important improvement to the above outlined typical path of an event generation. Namely, it employs self-adaptive Monte Carlo program \cite{Jadach-FOAM}. The event generation proceeds in two phases: exploration and generation. 

In the \textbf{Exploration phase}
\begin{itemize}
\item{Foam generates a vector of random variables $\mathcal{R}$ and feeds it to a few-body generator $\mathcal{G}$, chosen as the most appropriate for the considered problem,}
\item{$\mathcal{G}$ produces events which fulfil the energy-momentum conservation and corresponding phase space weight $\mathcal{W}_{PhS}$,}
\item{matrix element square is evaluated, providing the weight $\mathcal{W}_{ME}$ inside detector acceptance and set to zero outside,}
\item{product of weights $\mathcal{W}=\mathcal{W}_{PhS} \cdot \mathcal{W}_{ME}$ provides measure of the event density for self-adapting MC (Foam) in exploration phase.}
\end{itemize}
In the \textbf{Generation phase}
\begin{itemize}
\item Foam using the weight distribution found in the exploration phase provides appropriately weighted random vectors $\mathcal{R^{'}}$ 
that transformed into the final state particles four-momenta produce events corresponding to the assumed matrix element and the detector acceptance cuts,
\item if required, the events can be produced with an unit weight by use of rejection method.
\end{itemize}
In the next section we describe technical details of realization of this scheme in the C++ language, with emphasis on its object-oriented features.
\section{Program structure and class description}\label{description}
GenEx consists essentially of three main  components:
\begin{itemize}
\item
Generation of an event and estimation of its phase space weight, managed by \classMark{TEventMaker...} classes described in subsection \ref{event};
\item
Estimation of the matrix element and the corresponding event weight managed by \classMark{TWeight...} classes, described in subsection \ref{weight};
\item
Application of the acceptance cuts, managed by \classMark{TAcceptance} class described in subsection \ref{acceptance}.
\end{itemize}
In addition, a number of utility classes is provided to manage the generator initialization, cross-section estimation, histogramming and saving events. They are described in subsection \ref{utility}.
The adaptation of the density of points in the multidimensional space for the efficient event generation is realized
by Foam from ROOT library \cite{ROOT_site}. It requires a specification of \classMark{TDensity} class which inherits from \classMark{TFoamIntegrand} from ROOT package. The interaction between Foam and GenEx, described in general terms in Section \ref{general}, is explained by comments within the code of the generator main class \classMark{Generator}.

\subsection{Event generation: \classMark{TEventMaker}}\label{event}
\begin{figure}[htb]
\centering
 \includegraphics[height = 0.3\textheight, width = 1.0\textwidth]{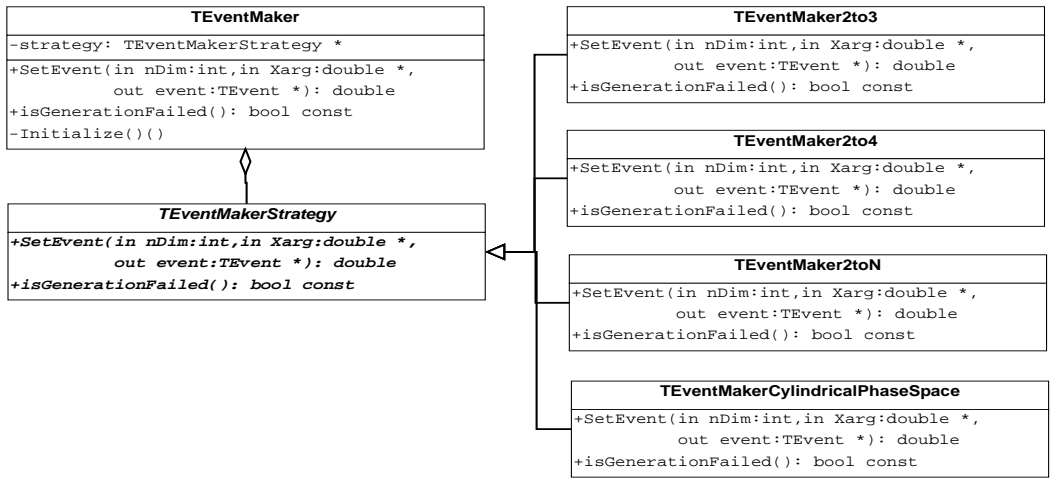}
\caption{The strategy pattern adapted to the generation of events from a set of random numbers. Abstract/Interface class \classMark{TEventMakerStrategy} is realized in different strategies: \classMark{TEventMakerCylindricalPhaseSpace}, \classMark{TEventMaker2to3}, \classMark{TEventMaker2to4} and \classMark{TEventMaker2toN}. The choice is made using the method \codeMark{Intialize()} in class \classMark{TEventMaker}.}
\label{Fig_EventMaker}
\end{figure}

The first functional part of the program converts a vector of random numbers (provided by Foam) into a set of 
generated particles four-momenta that preserves energy-momentum conservation. The generator contains 
several methods suitable for the generation of peripheral processes. The choice of the method is done in the 
framework of strategy pattern \cite{GangOfFourBook} (see Fig. \ref{Fig_EventMaker}), i.e., by implementation of the general interface/abstract class \classMark{TEventMakerStrategy}. Extraction of the interface as a separate class makes it also
possible to use all generator strategies in stand-alone programs (i.e. without Foam, useful when testing 
and debugging a new strategy). The different strategies represent different approaches to the phase space calculation. Owing the interface definition, the users can extend the generator by defining their own strategies. 
Each strategy provides the phase space part of the event weight $\mathcal{W}_{PhS}$ corresponding to the Lorentz invariant phase space (LIPS) defined \cite{Pilkuhn} as
\begin{equation}\label{n-body}
 \dif^{n}LIPS=(2\pi)^{4}\delta(p_{a}+p_{b}-\sum_{i=1}^{n}p_{i})\prod_{i=1}^{n}\frac{\dif^{3}p_{i}}{(2\pi)^{3}2E_{i}},
\end{equation}
where $n$ is the number of the final state particles, $p_{i}$ are their four-momenta, $p_{a}$ and $p_{b}$ are the beam four momenta and $E_{i}$ are the energies of the final state particles. The expression for the weight depends on the choice of the kinematic variables and the importance sampling, if applied. The strategies described below are characterized by a separation of the longitudinal and transverse components of momenta of some particles, which makes them suitable for integration of peripheral processes with exclusive particle production. 
The \classMark{TEventMaker} class contains the method \codeMark{Initialize()} that is responsible for choosing of strategy according to the prescription given in the configuration file \dirFileMark{Generator.dat}. 
\par
It should be noted that in all below described methods of the phase space generation it is assumed that particles $1$ and $2$ are peripherally scattered i.e. $p_{1,z} \cdot p_{a,z} > 0$ and $p_{2,z} \cdot p_{b,z} > 0$. The events with  $p_{1,z} \cdot p_{a,z} <  0$ or  $p_{2,z} \cdot p_{b,z} < 0$ belong to the second branch of the phase space (second root of the energy momentum conservation equation) and are suppressed in the generation (see Appendix). Below we shortly describe the strategies implemented in the present version of the generator.

\subsubsection{TEventMakerCylindricalPhaseSpace}\label{cylindrical}
\classMark{TEventMakerCylindricalPhaseSpace} class implements generation of the cylindrical phase space basing on the algorithm proposed in \cite{Jadach_cylindrical}. The transverse momenta and rapidities of the final state particles are chosen as the integration variables. The rapidities are generated uniformly in the phase space, for the transverse momenta importance sampling is employed. The latter method is useful whenever the transverse momenta of produced final state particles decrease sharply with increasing transverse momentum. In principle this method can be used for an arbitrary number of particles. The final state with $4$ particles implies the total dimension of probabilistic space in Foam initialization equal to $12$ and in GenEx this number is automatically derived and provided by \classMark{TPolicyReader}. The average transverse momentum of the generated particles is an adjustable parameter. It should be noted that the algorithm implemented here differs from the original one described in \cite{
Jadach_cylindrical} by suppression of events in which particles 1 and 2 change direction along z-axis, as mentioned 
above.  
\subsubsection{TEventMaker2to3}\label{3-body}
\classMark{TEventMaker2to3} class implements generation of $3$-body phase space using the variables proposed in \cite{Leb-Szcz-2}: 
\begin{itemize}
\item transverse momenta of scattered protons \\$p_{t1},p_{t2}$, $p_{min} \leq p_{t1},p_{t2} \leq p_{max}$;
\item azimuthal angles of the scattered protons $\phi_1$ and $\phi_2$;
\item rapidity of the intermediate particle $ y_{min} \leq y  \leq y_{max} $.
\end{itemize}
The phase space element can be expressed as
\begin{multline}\label{3-body-eq}
\dif^{3}LIPS(p_{1},p_{2},p_{3})=|\mathcal{J}|^{-1}(p_{1t}p_{2t},\phi_1,\phi_2,y)_{|root}\frac{1}{(2\pi)^{5}}\frac{1}{2E_{1}} \\
\frac{1}{2E_{2}}\frac{1}{2}p_{1t}p_{2t}dp_{1t}p_{2t}d\phi_1d\phi_2dy,
\end{multline}
where the transformation Jacobian $\mathcal{J}(p_{1t}p_{1t},\phi_1,\phi_2,y)_{|root}$ is calculated (see Appendix \ref{A}) at the root of the energy and longitudinal momentum conservation equations. In this case Foam is initialized with total dimension equals $5$. 
\subsubsection{TEventMaker2to4}\label{4-body}
\classMark{TEventMaker2to4} class implements generation of $4$-body phase space using the variables proposed in \cite{Leb-Szcz-1}:  
\begin{itemize}
\item transverse momenta of scattered protons $p_{t1},p_{t2}$, $p_{min} \leq p_{t1}, p_{t2} \leq p_{max}$,
\item protons azimuthal angles  $ 0 \leq \phi_{1},\phi_{2} \leq 2\pi $,
\item rapidities $y_1,y_2$ of the centrally produced particles $ y_{min} \leq  y_1,y_2  \leq y_{max} $,
\item length $p_{mt}$ and the azimuthal angle $\phi_{pmt}$ of the difference of the transverse momenta of the centrally produced particles, $0 \leq p_{mt} \leq p_{mt,max}$;
$0 \leq \phi_{pmt} \leq 2\pi$.
\end{itemize}
Using these variables phase space element can be expressed as
\begin{multline}
\dif^{4}LIPS(p_{1},p_{2},p_{3},p_{4})=|\mathcal{J}|^{-1}(p_{1t}, \phi_{1},p_{2t},\phi_{2},y_{3},y_{4},p_{mt},\phi_{m})_{|root}\frac{1}{(2\pi)^{8}}\frac{1}{2^{4}}\\
\frac{1}{2E_{1}}\frac{1}{2E_{2}}p_{1t}dp_{1t}d\phi_{1}p_{2t}dp_{2t}d\phi_{2}dy_{3}dy_{4}\dif^{2}p_{mt},
\end{multline}
where the transformation Jacobian $\mathcal{J}(p_{1t},p_{2t},\phi_1,\phi_2,y_1,y_2, p_{mt}, \phi_{pmt})_{|root}$ is calculated (see Appendix \ref{A}) at the root of energy and longitudinal momentum conservation equations. Foam is initialized with total dimension equals $8$. The range of generated variables has to be defined by the user. The functionality of this strategy is in principle the same as that of \classMark{TEventMaker2toN} with $N=4$, however, its choice may be preferential in situations when the constraint on rapidities of the centrally produced particles is required at the generator level.

\subsubsection{TEventMaker2toN}
\label{N-body}
\classMark{TEventMaker2toN} class implements generation of $N$-body phase space  using the recurrence relation \cite{Pilkuhn}
\begin{multline}\label{recurrence}
\dif^{N}LIPS(s;p_{1},p_{2},...,p_{N})=\frac{1}{2\pi}\dif^{3}LIPS(s;p_{1},p_{2},p_{3})\\\dif^{N-2}LIPS(M^{2};p_{4},p_{5}...p_N)dM^{2},  
\end{multline}
where $M$ is the invariant mass of an intermediate particle $3$ which subsequently decays into  particles $4\ldots N$.
In the first step the 3-body phase space is generated using variables described in \ref{3-body} and additionally in the variable $M$. In the second step, the centrally produced object decays according to the phase space into 
particles $4\ldots N$. This decay is managed by the class \classMark{TDecay} adapted from ROOT class of \classMark{TGenPhaseSpace}. In the final step the four-momenta of the decay products are boosted to the global rest frame. Thus the kinematic variables from which the final state four-momenta are constructed are:
\begin{itemize}
\item transverse momenta of peripherally scattered particles \\ $p_{t1},p_{t2}$, $p_{min} \leq p_{t} \leq p_{max}$,
\item azimuthal angles $\phi_1$ and $\phi_2$ of particles  1 and 2: $ 0 \leq \phi_{1,2} \leq 2\pi $,
\item rapidity of the intermediate object $ y_{min} \leq y \leq y_{max} $,
\item mass $M$ of the intermediate object  $ M_{min} \leq M \leq M_{max} $,
\item remaining $3N-10$ variables are managed by \classMark{TDecay}.
\end{itemize}
For $N=4$ functionality of this strategy is in principle the same as that of \classMark{TEventMaker2to4}, however, its choice may be preferential in situations in which constraint on the invariant mass $M$ of the centrally produced system is required at the generator level.
The code of \classMark{TDecay} class which manages the decays of the centrally produced object results from 
adaptation of ROOT utility class \classMark{TGenPhaseSpace}. The adapted features are the following:
\begin{itemize}
\item
random numbers are provided by Foam,
\item
the decay phase space weight $\mathcal{W}_{decay}$ of the centrally produced object many-body decay corresponds to formula (\ref{recurrence}).
\end{itemize}
In addition to \codeMark{getWeight()} method, which provides phase space weight of the event, two additional methods are provided:
\begin{itemize}
\item \codeMark{GetDecayWeight()} - returns the phase space weight  $\mathcal{W}_{decay}$ of the decay $M \rightarrow 3+4...+N $,
\item \codeMark{GetDecayPhaseSpaceIntegral()} - returns \\ $Idecay=LIPS(M,n,m_1,m_2,...m_n)$ of the integrated decay phase space of the central object as a function of its mass $M$. This method will work only if \configOptionMark{TEventMaker2toN::bIDecay=1} (see subsection \ref{options} in which other options for \classMark{TEventMaker2toN} are defined). 
\end{itemize}
The normalized decay weight $\mathcal{W}_{decay}/Idecay$ represents the probability of a particular configuration of particles from the decay in a given event (original \classMark{TGenPhasespace} provides only the relative probabilities). It can be useful when the matrix element represents peripheral production of the resonance.

The class \classMark{TDecayIntegral} serves as a tool for calculation of the value of the phase space integral for decay into a given final state as a function of the system invariant mass $M$. In the constructor of \classMark{TEventMaker2toN} table of values of the integral in bins of $M$ is calculated. The method \codeMark{GetDecayPhaseSpaceIntegral()} employs a linear interpolation to calculate the integral value for a given mass $M$.   

\subsection{Event weight: \classMark{TWeight}}\label{weight}
\begin{figure}[htb!]
\centering
 \includegraphics[height = 0.3\textheight, width = 1.0\textwidth]{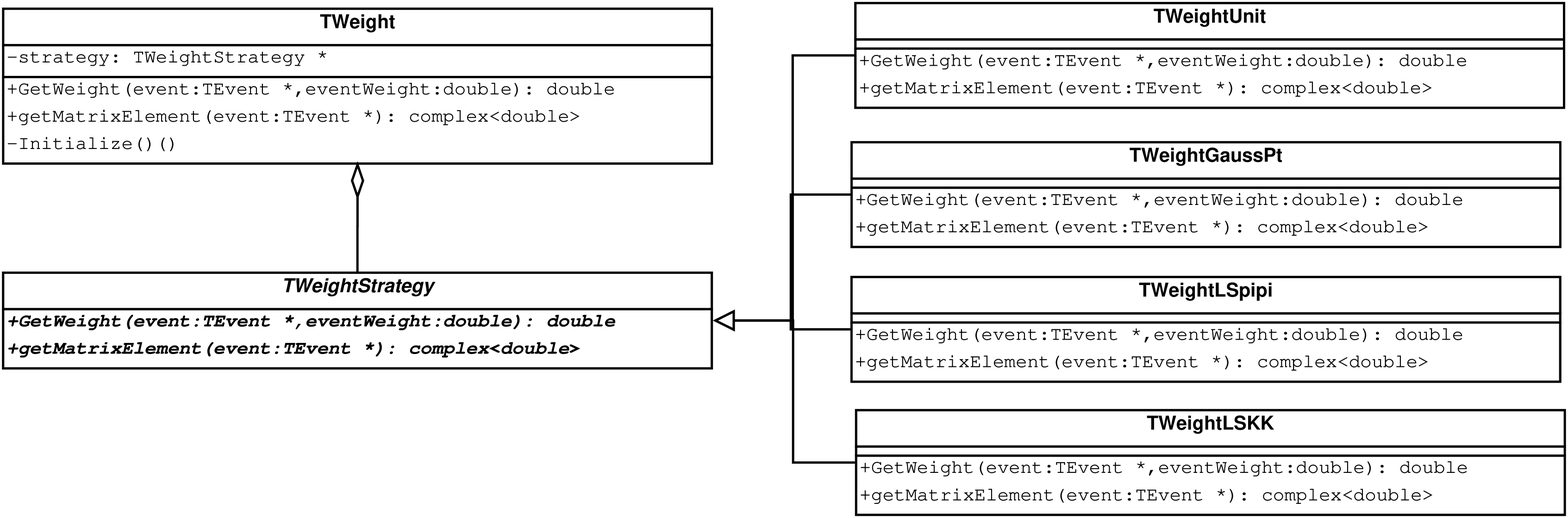}
\caption{The strategy pattern adapted to calculate weight and matrix element for the  event. Abstract/Interface class \classMark{TWeightStrategy} is realized as \classMark{TWeightUnity}, \classMark{TWeightGaussPt}, \classMark{TWeightLSpipi} or \classMark{TWeightLSKK} using the method \codeMark{Intialize()}.}
\label{Fig_Weight}
\end{figure}

Once the event is generated and stored in \classMark{TEvent} object its weight depending on final state particles four-momenta can be calculated. The phase space weight $\mathcal{W}_{PhS}$ is already calculated by \classMark{TEventMaker}. The module which calculates the value of the matrix element and its square $\mathcal{W}_{ME}$ is implemented as a strategy pattern, thus allowing for a choice of the model (see Fig. \ref{Fig_Weight}). The binary ($0/1$) acceptance weight is calculated in \classMark{TAcceptance} class described in the next subsection. 
The interface class  from which all particular  strategies inherit is called \classMark{TWeightStrategy}. The classes which implement this interface include  two methods important for the user 
\begin{verbatim}
 double GetWeight( TEvent * event, double eventWeight )
\end{verbatim}
which calculates the total weight for a given event, i.e., the quotient of the phase space weight (including Jacobian), the matrix element square and the acceptance. The second important method is 
\begin{verbatim}
 complex<double> getMatrixElement( TEvent * event )
\end{verbatim}
which returns the matrix element value for an event. The following strategies are implemented

\begin{itemize}
 \item{\classMark{TWeightUnit} - unit weight (for testing),}
 \item{\classMark{TWeightGaussPt} - Gaussian distribution of the transverse momenta of all outgoing particles,}
 \item{\classMark{TWeightLSpipi} - matrix element for process $pp\to pp \pi^{+}\pi^{-}$ \cite{Leb-Szcz-1},}
 \item{\classMark{TWeightLSKK} - matrix element for process $pp\to pp K^{+}K^{-}$ \cite{Leb-Szcz-3}.}
\end{itemize}
The \codeMark{Initialize()} method allocates appropriate object that realizes a given weight generation strategy during the initialization according to the data input from the configuration file. If the user wants to add a new weight/matrix element to the generator then the implementation of \classMark{TWeightStrategy} interface should be created and a corresponding change in \codeMark{Initialize()} arranged.

\subsection{Acceptance: \classMark{TAcceptance}}\label{acceptance}
\begin{figure}[htb!]
\centering
 \includegraphics[height = 0.3\textheight, width = 1.0\textwidth]{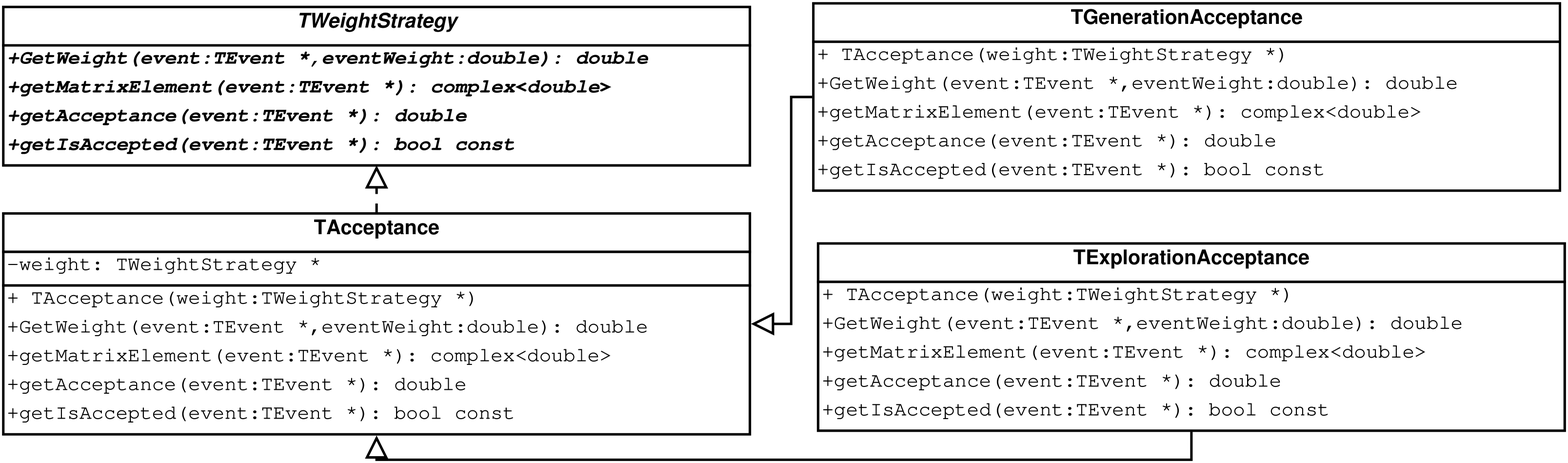}
\caption{The decorator pattern adapted to decorate \classMark{TWeight} class. Class \classMark{TAcceptance} implements \classMark{TWeightStrategy} interface and also contains pointer to concrete realization of this interface. \classMark{TAcceptance} is inherited by two classes: \classMark{TExplorationAcceptnace}, which is used during the exploration phase and \classMark{TGenerationAcceptance} used in generation of events. }
\label{Fig_Acceptance}
\end{figure}

The acceptance cuts on kinematic variables other than those directly generated should be applied after the object \classMark{TEvent} is created and before the matrix element is calculated. Therefore, it is natural to implement the acceptance as a decorator pattern \cite{GangOfFourBook} that decorates \classMark{TWeight} class, see Fig. \ref{Fig_Acceptance} for details. 
The \classMark{TAcceptance} class has the same interface as the \classMark{TWeight} class, namely \classMark{TWeightStrategy}. The constructor of the acceptance class takes as an argument a pointer to the object that realizes \classMark{TWeightStrategy} interface, which will be decorated. The weight and the matrix element of the event can be calculated in \classMark{TAcceptance}  calling \codeMark{GetWeight(...)} and \codeMark{getMatrixElement(...)} methods. If the acceptance test result is positive the event is passed to the internal class that calculates the matrix element. If the event does not pass the acceptance test, zero is returned without calling internal decorated \classMark{TWeightStrategy} object. In this way calculation of the matrix element outside the acceptance region is avoided. Two additional methods
\begin{verbatim}
 double getAcceptance( TEvent * event );
 bool getIsAccepted( TEvent * event );
\end{verbatim}
return double ($0.0/1.0$) and boolean value respectively according to the acceptance of the event.

There are two classes derived from \classMark{TAcceptance}: \classMark{TExplorationAcceptnace} and \classMark{TGenerationAcceptance}. Both of them contain a set of different cuts that can be chosen during the creation of an object according to the data in  the configuration file. The second class contains the cuts that are applied during the generation phase. Sometimes, however, there is a need to perform the Foam exploration in a wider phase space (with less restrictive cuts), e.g., see subsection \ref{exploration}, and for this purpose \classMark{TExplorationAcceptnace} class was created. 
Each of the derived acceptance classes contains the set of functions that realize the cuts under the names \codeMark{cutX()}, where $X$ is some nonnegative integer number. The cut functions are included during the compilation into \classMark{T...Acceptance} classes from the files in \dirFileMark{ACCEPTANCE\_CUTS} directory.

\subsection{Utility modules}\label{utility} 
\subsubsection{Initialization}\label{configuration}
Initialization of the generator configuration i.e. of the set of parameters and numeric constants which define chosen generator  options, generated process and matrix element (reaction model) is managed by \classMark{TConfigurationReader} and \classMark{TPolicyReader} classes. \classMark{TConfigReader} serves as the configuration bank. It keeps the configuration values 
in a simple key-value map form. It also contains a parser which reads the configuration from a text file with the structure: 
\begin{verbatim}
variable_name1 = value1 #comment1
...
#comment
...
variable_name2 = value2 ;comment2
...
\end{verbatim}
The comments start from $\#$ or $;$ and end with a new line. The values can be extracted from the object of 
the class by a set of the methods \codeMark{getTYPEValue()}, where TYPE is the type of the value, e.g., 
\codeMark{getDoubleValue()}, \codeMark{getStringValue()}, etc. Such method gets a string, which is a name of a variable in the file. In case such a variable is not defined in the file it throws an exception. The class \classMark{TConfigReader}
allows reading of the data from the configuration file. Therefore, it is necessary to introduce another class that
can impose constraints on the parameters and calculate derived quantities. This class, \classMark{TPolicyReader},
which inherits from \classMark{TConfigReader}, reads the initial configuration file (\dirFileMark{Generator.dat}) and interprets it
as follows:
\begin{enumerate}
\item
the dimension of the phase space \configOptionMark{kDim} is set according to \configOptionMark{eventGenerationStarategy} configuration variable,
\item
the centre of mass energy \configOptionMark{tecm} and the boost to laboratory frame are calculated from the momenta of initial 
particles,
\item the configuration file of model $X$,  i.e.\dirFileMark{/MODEL\_DATA/X.dat}, is read and model constants are added to the list of all previously defined constants.
\end{enumerate}
\subsubsection{Histograming}\label{histograming}
The \classMark{THistogram} class provides methods to calculate various kinematic variables and to fill, draw and save histograms in the PostScript and root files. The method \codeMark{WriteHistograms( double xsection )} that draws and saves histograms to disk, called in the final part of the main class \classMark{Generator}, allows the user to normalize selected distributions as the differential cross-section histograms. The \classMark{THistogram} class using a built-in strategy pattern with the interface class \classMark{THistogramStrategy} provides separate methods to calculate the kinematics and to fill histograms for different numbers of particles in the final state. \\
\subsubsection{Integration}
The \classMark{TIntegral} class calculates the value of the Monte Carlo integral (cross-section) and its statistical uncertainty  using the average weight and the standard deviation of weight distribution respectively. If the adaptive Monte Carlo integrator has been switched off by setting \configOptionMark{IntegratorSetup=2} this class is used to calculate the integral and its error. Otherwise, integration is performed within Foam. The class is also used to count the number of events that are accepted when using different exploration and generation cuts.\\
\subsubsection{Logging}
The \classMark{TLog} class provides logging of two types: global and configuration. The global logging is realized by the singleton instance \cite{GangOfFourBook} of \classMark{TLog} class contained in \dirFileMark{Global.h} header file, and produces \configOptionMark{RunLogFile}, which contains basic information for a given generator run (time/date, initialization of Foam, initialization of Generator and errors, if any). The configuration logging saves the configuration data into a text file, which name is set up by \configOptionMark{ConfigLogFile} parameter, \dirFileMark{ConfigLog.log} by default. It contains all the relevant constants and parameters from \dirFileMark{Generator.dat} and \dirFileMark{/MODEL\_DATA}. The \dirFileMark{ConfigLog.log}, alphabetically sorted by the Python script \dirFileMark{sortConfigLog.py} configuration backup, 
can be archived and used to run the generator by the \commandMark{make test} command, as described in subsection \ref{example}.

Note that every utility module mentioned above is designed in such a way that it can be used independently in a user program. These modules can be used to perform additional analyses of the data or when constructing a new generator or weight strategy.
\subsection{Simplified sequence diagrams}\label{sequence}
In this section a simplified UML sequence diagrams for the generator are presented. They visualize interaction between different parts of the generator and the sequence of calls. The main program in \dirFileMark{main.cxx} file creates instance of the  \classMark{Generator} class. This object in constructor creates all other objects needed in generation. Now, the program performance path splits depending on set-up parameter \configOptionMark{IntegratorSetup = 1 or 2}.
\par
\begin{figure}[htb!]
\centering
 \includegraphics[height = 0.4\textheight, width = 1.0\textwidth]{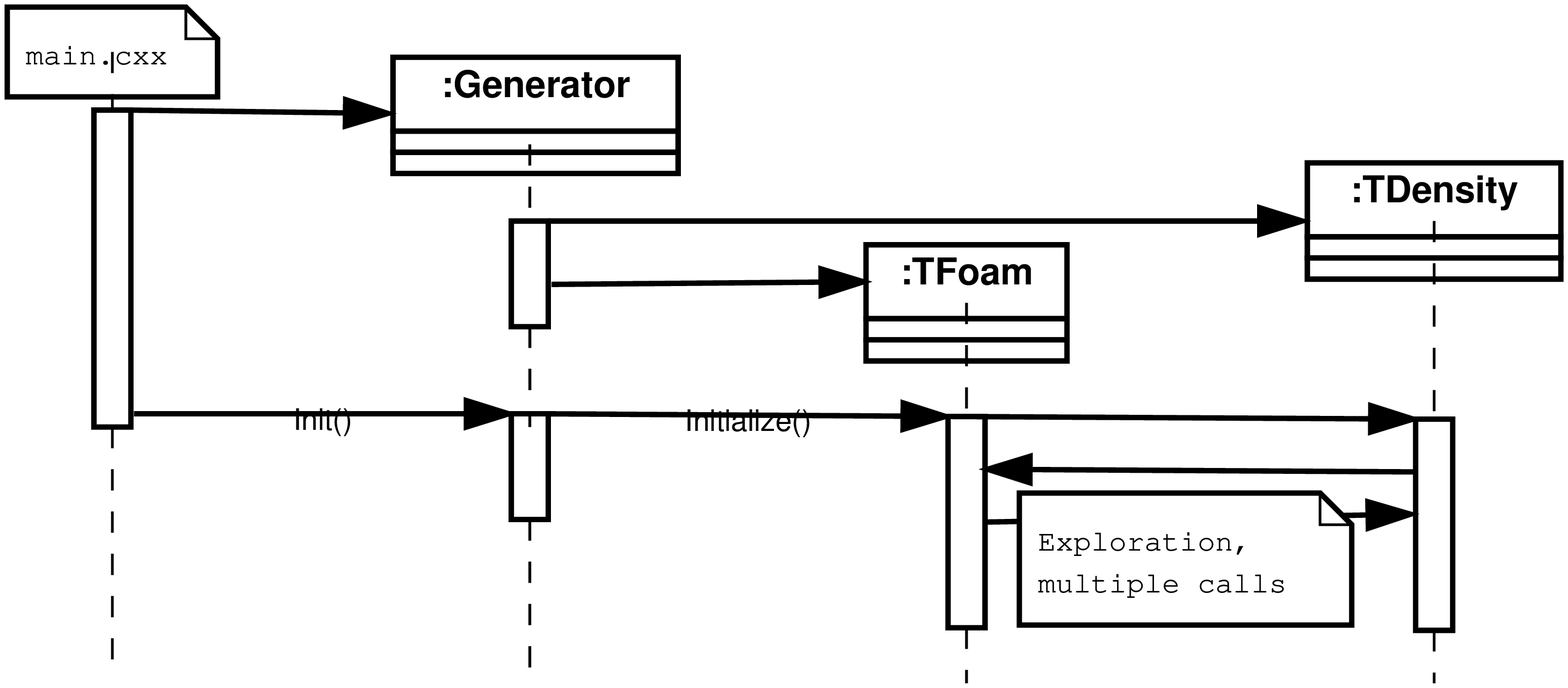}
\caption{Exploration phase using Foam. \classMark{TDensity} class internal calls structure was not presented.}
\label{Fig_FoamExploration}
\end{figure}

\begin{figure}[htb!]
\centering
 \includegraphics[height = 0.4\textheight, width = 1.0\textwidth]{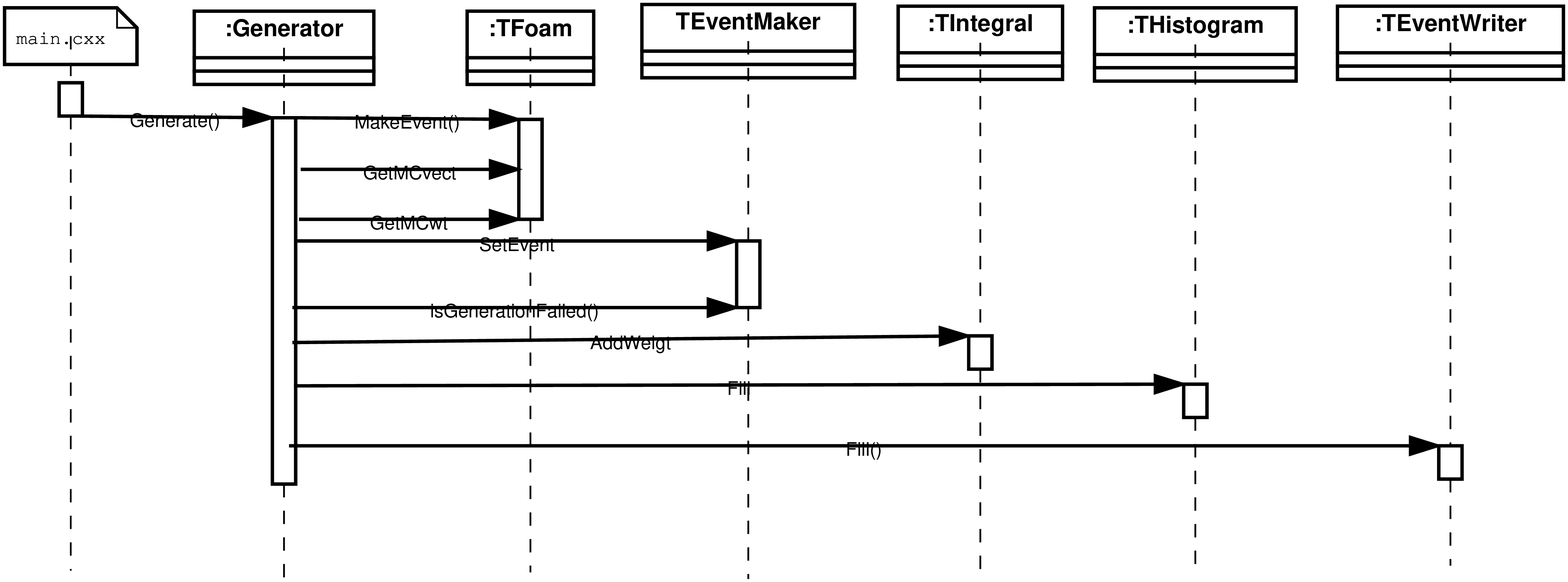}
\caption{Generation phase when when self-adapting Monte Carlo method Foam is used. \classMark{TFoam} object internally calls \classMark{TDensity} object which was not shown here to keep diagram simple.}
\label{Fig_FoamGeneration}
\end{figure}

 
In the self-adapting Monte Carlo mode (\configOptionMark{IntegratorSetup = 1}) Foam is initialized and exploration phase starts. This part is presented in form of sequence diagram, Fig. \ref{Fig_FoamExploration}. After successful exploration phase the  generation starts. The generation sequence diagram is presented in Fig. \ref{Fig_FoamGeneration}.
\par
\begin{figure}[htb!]
\centering
 \includegraphics[height = 0.4\textheight, width = 1.0\textwidth]{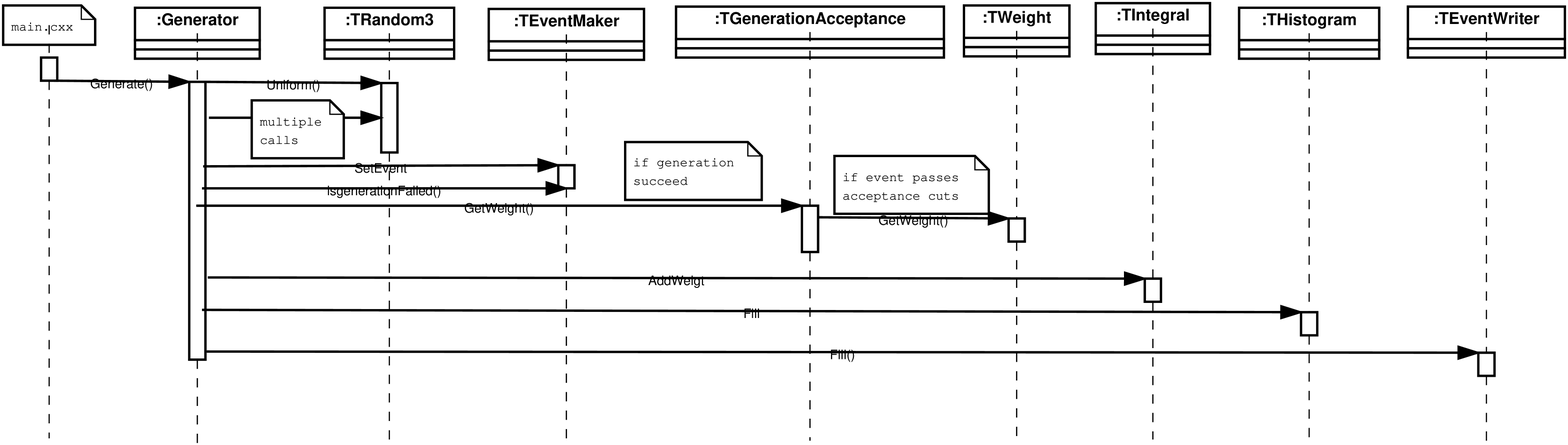}
\caption{Generation phase when when self-adapting Monte Carlo method Foam  is not used.}
\label{Fig_RandomGeneration}
\end{figure}

 
The second path of the program performance corresponding to \configOptionMark{IntegratorSetup = 2} is presented in Fig. \ref{Fig_RandomGeneration}.
Here instead of Foam an object of \classMark{TRandom3} provides a vector of random numbers uniformely distributed on the unit interval.
\section{How to install and use GenEx}
\subsection{Installation, Makefile and utility scripts}\label{instalation}

Prerequisites for compiling and running of the program are as follows
\begin{itemize}
 \item {ROOT \cite{ROOT_site}, }
 \item {GNU C++ compiler \cite{GNUCpp},}
 \item {GNU Makefile \cite{GNUMAKE},}
 \item {Doxygen - for the documentation generation \cite{Doxygen} (optional),}
 \item {Python interpreter - for running additional scripts \cite{Python} (optional).}
\end{itemize}
The code of the GenEx generator in a form of a compact file \dirFileMark{GENERATOR.tar.gz}  can be obtained from the authors on request. 
The command \commandMark{tar -zxf GENERATOR.tar.gz} unpacks it to \dirFileMark{GENERATOR} directory, which has the following content:
\begin{itemize}
 \item {\dirFileMark{GENERATOR/} - main directory contains a simple GNU Makefile \cite{GNUMAKE} and scripts that perform various operations,}
 \item {\dirFileMark{GENERATOR/bin} - contains \dirFileMark{main.x} generator program that appears after the compilation;}
 \item {\dirFileMark{GENERATOR/include} - contains the header files,}
 \item {\dirFileMark{GENERATOR/src} - contains source files and the main generator file \dirFileMark{main.cxx},}
 \item {\dirFileMark{GENERATOR/build} - is the temporary directory for building. It is created by \commandMark{make run} command and deleted with all the content by \commandMark{make clean} command,}
 \item {\dirFileMark{GENERATOR/EVENTS} - contains the files with stored events from the run,}
 \item {\dirFileMark{GENERATOR/MODEL\_DATA} - contains the configuration (text) files for different models (e.g. matrix elements) used in generation,}
 \item {\dirFileMark{GENERATOR/ACCEPTANCE\_CUTS} - contains the files defining cuts,}
 \item {\dirFileMark{GENERATOR/Tests} - contains the sorted generator configuration files like \dirFileMark{ConfigLog.log} for various tests (see \commandMark{make test} below).}
\end{itemize}
\dirFileMark{Makefile} assumes that \codeMark{\$ROOTSYS} and \codeMark{\$ROOTINC } environment variables are properly set to  ROOT and ROOT/include directories.
\commandMark{make} command has the following options: 
\begin{itemize}
\item \commandMark{make all} - compiles the generator,
\item \commandMark{make run} - compiles and runs the generator using the configuration file \dirFileMark{Generator.dat}; the file \dirFileMark{ConfigLog.log} that contains configuration of the run is also created and can be archived, edited and used for running the generator by means of \commandMark{make test} tool,
\item \commandMark{make test} compiles and runs the generator using the short and sorted configuration file \dirFileMark{ConfigLog.log}. The command prompts the user to provide the sequential number of configuration file stored in \dirFileMark{/Tests} directory and runs the generator. Examples of usage of \commandMark{make test} command are described in subsection \ref{example},
\item \commandMark{make clean} - restores \dirFileMark{Genex} directory to the original state, i.e., removes all the files created in the process of compiling and linking of the program; data like histograms, events and log informations are unaffected,
\item \commandMark{make line-count} - counts the lines of code of the program,
\item \commandMark{make Generate-doc} - generates the Doxygen documentation.
\end{itemize}
The commands \commandMark{make run} and \commandMark{make test} will produce an output with result of the Monte-Carlo integration, 
the file \dirFileMark{histograms.ps} with default histograms and the file \dirFileMark{histograms.root} which can be browsed typing: \commandMark{root browseHistograms.C}
\subsection{Setting parameters}
All parameters which define the options of the generator, the generated process and names of output files are set in the main 
configuration file \dirFileMark{Generator.dat}, which has format described in subsection \ref{configuration}. The constants of the implemented models of exclusive processes are set in the separate data files in the directory \dirFileMark{MODEL\_DATA}.  
\subsubsection{General set-up}\label{gen-setup} 
This group of parameters defines general set-up of the program - method of the phase space generation and  output of the generator:
\begin{itemize}
\item
\configOptionMark{NumberOfEventsToGenerate} - total number of events to be generated in the generation phase. These events are either weighted or unweighted 
depending on Foam parameter \configOptionMark{OptRej} $0/1$;
\item
\configOptionMark{IntegratorSetup} - choice between the adaptive Monte Carlo Foam integrator and the standard Monte Carlo integration. The second choice is
usefull for  testing a new generator or weight strategy;
\item
\configOptionMark{seed} - random number generator seed. This variable can be used to run parallel programs with different seeds to generate statistically
 independent sets of data; 
\item
\configOptionMark{Foam\_parameters} - group of steering parameters for Foam. Their meaning is explained in \cite{Jadach-FOAM}. The role of two of them (\configOptionMark{nCells}, \configOptionMark{nSampl}) is explained in the next subsection;
\item
\configOptionMark{SaveEventDataLHE\_XML}, \configOptionMark{SaveEventDataLHE\_TXT}, \configOptionMark{SaveEventDataRoot} - switch on/off writing generated events to file in a chosen format (LHE-XML, LHE-TXT or ROOT), see \cite{LesHouchesAccords1,LesHouchesAccords2} for LHE format specification. Templates to read these files are contained in the folder \dirFileMark{Analysis};
\item
\configOptionMark{RootEventFile}, \configOptionMark{LHEEventFile}, \configOptionMark{LHETxtEventFile} - set names of event backup files. These files can be analysed using additional programs in \dirFileMark{Analysis} folder described in subsection \ref{analysis}; 
\item
\configOptionMark{eventGeneratorStrategy} - choice between  strategies of event generation  described in subsection \ref{event}:
\begin{itemize}
\item
\#1 - cylindrical phase space described in subsection \ref{cylindrical},
\item
\#2 - 3-body phase space described in \ref{3-body},
\item
\#3 - 4-body phase space described in \ref{4-body},
\item 
\#4 - N-body phase space described in \ref{N-body}.
\end{itemize}
\end{itemize}
\subsubsection{Specification of the initial and final state}\label{fin-setup}
\begin{itemize}
\item Parameters defining the initial state particles and frame of reference:
\begin{itemize}
\item[-]\configOptionMark{frametype} - type of the reference frame: centre of mass (CM) or laboratory (LAB), 
\item[-]\configOptionMark{tecm} - CM energy of the collision; has to be specified if \configOptionMark{frametype=CM},
\item[-]\configOptionMark{idA, idB} - PDG names of the initial particles,
\item[-]\configOptionMark{EA, EB} - energies of colliding particles, have to be specified if \configOptionMark{frametype=LAB}.
\end{itemize}
\item Parameters defining the final state particles:
\begin{itemize}
\item[-]\configOptionMark{X::massIndication} = $0$ if particle mass is specified in configuration file, =$1$ if particle mass is derived from PDG code, where 'X' denotes the event maker type e.g. \classMark{TEventMaker2toN};
\item[-]\configOptionMark{X::id1,X::id2...X::idN} - PDG names of particles in the generated process, where 'X' denotes the event maker type e.g. \classMark{TEventMaker2toN};
\item[-]\configOptionMark{X::m1,X::m2...X::mN} - masses of particles in the generated process, where 'X' denotes the event maker type e.g. \classMark{TEventMaker2toN}.
\end{itemize}
\end{itemize}
It should be noted that particles $1$ and $2$ have been singled out of the rest as peripherally scattered for reasons explained in Section \ref{event}.
\subsubsection{Parameters and options for phase space generation}\label{options}
\begin{itemize}
\item Parameters for generation of 3-body phase space described in subsection \ref{3-body}: 
\configOptionMark{X::p\_max,X::p\_min,X::y\_min,X::y\_max}, where X stands for \classMark{TEventMaker2to3}.
\item Parameters for generation of 4-body phase space described in subsection \ref{4-body}: \configOptionMark{X::p\_max,X::p\_min,X::y\_min,X::y\_max,X::pmt\_max}, where X stands for \classMark{TEventMaker2to4}.
\item Parameters for generation of N-body phase space described in subsection \ref{N-body}: 
\begin{itemize}
\item[-] \configOptionMark{centralMassOption} option defines of the central system mass range: $0$ - kinematic boundaries; $1$ - specific mass range $[\mathbf{m_{min},m_{max}}]$; $2$ - from threshold to $\mathbf{m_{max}}$
\item[-] ranges of kinematic variables \configOptionMark{X::p\_max,X::p\_min}\\,\configOptionMark{X::y\_min,X::y\_max} where X stands for \classMark{TEventMaker2toN}.
\end{itemize}
\end{itemize}
\subsubsection{Choice of the reaction model (matrix element)}
\configOptionMark{weightStrategy} - selects different existing weight strategies, i.e., class calculating matrix element for given process. Each class has its own configuration file which contains the model parameters:
\begin{itemize}
\item $1$ - unit weight,
\item $2$ - Gaussian distribution in transverse momenta of all outgoing particles,
\item $3$ - model for exclusive production of $\pi^+\pi^-$ pairs in the process $pp \rightarrow p\pi^+\pi^-p$ as described in \cite{Leb-Szcz-1},
\item $4$ - model for exclusive production of $K^+K^-$ pairs in the process $pp \rightarrow pK^+K^-p$ as described in \cite{Leb-Szcz-3}.
\end{itemize}
\subsection{Tool for analysis of event files in root and LHE formats}\label{analysis}
The event file in root or LHE formats can be analysed using additional programs contained in \dirFileMark{Analysis} directory. One has to perform the following steps:
\begin{itemize}
 \item {copy \dirFileMark{ConfigLog.log, integral.log} and one of the event files \dirFileMark{event.root, event.txt}  into  the 
 directory with appropriate analysis program,}
 \item {adjust \dirFileMark{analysis.cxx} program to the user needs, following the comments in the file,}
 \item {type \commandMark{make run} command to start the analysis.}
\end{itemize}
\subsection{How to use generator: quick start}\label{example}
After unpacking the generator from \dirFileMark{GENERATOR.tar.gz} and entering the \dirFileMark{GENERATOR} directory, the \commandMark{make run} will start generation of $20000$ events of the process $pp \rightarrow pp\pi^+\pi^-$ at $\sqrt{s} = 200$ $GeV$ collision energy using the matrix element of \cite{Leb-Szcz-1}. The program output to the screen contains information about \dirFileMark{TFoam} set-up parameters (see subsection \ref{gen-setup}) the results of the Foam phase space exploration phase (see section \ref{general}) and final Foam results, in particular the most precise cross section estimate $0.01365 \pm 3\cdot 10^{-5}$ $mb$. At this point it will be useful to inspect the files created by the program and observe effects of modifications of the main input file \dirFileMark{Generator.dat}. In the main generator directory \dirFileMark{GENERATOR} the following files are created:
\begin{itemize}
\item \dirFileMark{histograms.root} and  \dirFileMark{histograms.ps} contain root histograms and their PostScript representation created by \dirFileMark{THistogram4} class directly in the generator loop. This is the quickest method for analysis of the generated events during the model tuning or development of a new generator strategy;
\item \dirFileMark{events.root},\dirFileMark{events.lhe}, \dirFileMark{events.txt} from the \dirFileMark{EVENTS} directory are files in which generated events are stored. The tools provided in the directory \dirFileMark{Analysis} (see Section \ref{analysis}) enable user to develop analysis of the events without repeating generation.  
\item \dirFileMark{integral.log} contains the value of the final estimation of the MC integral (cross section) and its uncertainty. This file transferred to \dirFileMark{Analysis} directory is read by analysis.cxx tool \ref{analysis};
\item \dirFileMark{ConfigLog.log} records the generator set-up, including parameters contained in the \dirFileMark{Generator.dat} as well as the model parameters from the directory \dirFileMark{./MODEL\_DATA} used during the run. It can be archived, edited and used for re-run the generator as follows:
\begin{enumerate}
\item{copy \dirFileMark{ConfigLog.log} into \dirFileMark{Tests} directory and rename it to \dirFileMark{testX}, where $X$ is some integer number,}
\item {in Python script \dirFileMark{test.py} add appropriate entry into dictionary structure \codeMark{tests}, e.g., if the renamed file is \dirFileMark{test30}, then in the dictionary appropriate entry \codeMark{30: 'Description of the test'} should be added. Correct indentation should be used as in every Python script.}
\item{The command \commandMark{make test} will run the generator upon appropriate test number entry from the keyboard.}
\end{enumerate}
\item In the original \dirFileMark{Generator.dat} option \configOptionMark{useExplorationCuts} has been set to zero, so that identical acceptance cuts are applied both in exploration and generation phases (see sections \ref{general} and \ref{acceptance}). After setting \configOptionMark{useExplorationCuts=1} \commandMark{make run} will produce additional \dirFileMark{genhistos.ps} file, which contains histograms made using finer cuts.  
\end{itemize}
A new generator set-up requires the following actions:
\begin{itemize}
\item[-] edit parameter and options in \dirFileMark{Generator.dat} described in section \ref{gen-setup},
\item[-] if needed change the model-$X$ parameters in \dirFileMark{/MODEL\_DATA/X.dat},
\item[-] define the user acceptance cuts editing \\
\dirFileMark{/ACCEPTANCE\_CUTS/double TGenerationAcceptance::cut0( void )} and\\ 
\dirFileMark{/ACCEPTANCE\_CUTS/double TExplorationAcceptance::cut0( void )} or any other available cut functions. The circumstances in which  different exploration and generation cuts make
sense are explained in section \ref{exploration}.
\end{itemize}
In order to analyze the generation results a user can either edit \dirFileMark{THistogram3, THistogam4} or \dirFileMark{THistogramN} (depending on the number of particles in the final state, section \ref{histograming}) or to create analysis program in the directory \dirFileMark{Analysis} (see section \ref{analysis}). 
\subsection{Tests of the generator}
The generator was extensively tested for consistency of results obtained using different methods of phase space generation and for stability of results with respect to parameters in the generator set-up. The obtained results were compared to those published in  \cite{Leb-Szcz-1} and \cite{Leb-Szcz-3}. In every test $N= 10^{6}$ events was generated. All tests were performed using $nCells = nSampl = 10000$ during the exploration if not indicated otherwise. The generator configurations for all 26 tests are archived in the directory \dirFileMark{/Test}.
\subsubsection{Comparison of GenEx with published results}
In the Tab. \ref{Table-1} the cross sections for the process $pp \rightarrow pp\pi^+\pi^-$ at $200 ~\GeV$ obtained using \classMark{TEventMaker2to4} and \classMark{TWeightStrategyLS} (generator configuration corresponding to \dirFileMark{/Tests/test1... test3}) are compared to the calculations published in \cite{Leb-Szcz-1}. Similar tests have been performed for energies $500$ GeV, $7$ TeV and process $pp \rightarrow ppK^+K^-$ (\dirFileMark{/Tests/test4... test14}). Agreement between \cite{Leb-Szcz-1}, \cite{Leb-Szcz-3} and GenEx results is very good for all the tested energies and processes. 
\begin{table}[h]
\centering
\begin{tabular}{|c|c|c|c|}
\hline
Acceptance cuts & $\Lambda_{off}^2$ & ref. \cite{Leb-Szcz-1} & GenEx \\ [0.5ex]
\hline\hline
$p_{t} > 0.15 $;$\vert \eta \vert \leq 1$ & $1.0$ & $14.85$ & $14.90\pm 0.01$\\
\hline
$p_{t} > 0.15$;$\vert \eta \vert \leq 1$,$0.003 \leq -t_1,t_2 \leq 0.035 $
 & $1.0$ & $0.79$ & $0.79 \pm 0.0007$\\
\hline
$p_{t} > 0.15$$; \vert \eta \vert \leq1$,$0.003 \leq -t_1,t_2 \leq 0.035 $
 & $1.6$ & $1.53$ & $1.52 \pm 0.0003$\\
\hline
\end{tabular}
\caption{Comparison of GenEx with results of \cite{Leb-Szcz-1} for different acceptance cuts and model parameter $\Lambda_{off}^2$. Corresponding cross-section values in the last two columns are given in $\mathrm{\mu b}$, and $\Lambda_{off}^2$ in $GeV^2$.}
\label{Table-1}
\end{table}
\subsubsection{Stability of results with respect to generator parameters set-up}\label{stability}
The generator is not a magic box, which produces right output. To obtain correct results the user is requested to check if the number of cells (\configOptionMark{nCells}) and the number of samplings per cell (\configOptionMark{nSampl}) are sufficient for a given generator set-up. This is illustrated in Tab. \ref{Table-2} (\dirFileMark{/Tests/test22,...test26}). 
\begin{table}[ht]
\centering
\begin{tabular}{|c|c|c|c|}
\hline
\configOptionMark{nCells} & \configOptionMark{nSampl} & y-range&cross section $[\mikrob]$  \\
\hline\hline
 $1000$ & $1000$ & $[-8.0,8.0]$ &$0.977 \pm 0.002$ \\
\hline
$10000$ & $1000$ & $[-8.0,8.0]$&$0.977 \pm 0.002$ \\
\hline
$10000$ & $10000$ & $[-8.0,8.0]$&$1.421 \pm 0.004$ \\
\hline
$10000$ & $1000$ & $[-2.0,2.0]$&$1.519 \pm 0.004$ \\
\hline
$10000$ & $10000$ & $[-2.0,2.0]$& $1.523 \pm 0.004$\\
\hline
\end{tabular}
\caption{Cross-section calculation corresponding to the generator set-up indicated in the third row of Tab. \ref{Table-1}, obtained for different rapidity ranges at the generator level and different Foam parameters (\configOptionMark{nCells, nSampl}).}
\label{Table-2}
\end{table}
The generator set-up for rather restrictive acceptance cuts is run for different number of cells, samplings and rapidity
ranges of centrally produced particles. When the rapidity range is much wider, for example $[-8.0,8.0]$, that the one required by 
acceptance cuts $[-1.0,1.0]$ then $1000$ cells is not sufficient to obtain a correct result. Using $1000$ cells
and $1000$ samplings/cell one gets a result (first row of Tab. \ref{Table-2}) which differs by about $36$\% from the correct
result (last row of Tab. \ref{Table-2}) obtained with $10000$ and $10000$ of cells and samplings/cell respectively and smaller rapidity range. It should be noted that the error calculated by Foam is almost an order of magnitude smaller. However, when 
the rapidity range is set correctly (i.e. closer to that required by the acceptance cuts) $1000$ cells and $1000$ sampling/cell is completely sufficient to obtain correct (within the statistical uncertainty indicated by Foam) result.
\par
This example shows that GenEx is a tool which should be used with care. If the phase space volume within the acceptance cuts is significantly smaller  than the phase space in which the events are generated, then for a small number of samples (i.e. $Cells\cdot Samples$) there is no way to probe sufficiently the shape of the integrand. Thus, in the case of very restrictive acceptance cuts, the stability of results against Foam and the generator parameters set-up should be checked. One should note that sometimes instead of increasing the number of cells and sampling, the use of different exploration and generation cuts can lead to correct results (see subsection \ref{exploration}).
\subsubsection{Consistency between TEventMaker2to4 and TEventMaker2toN, N=4}
In this test (\dirFileMark{/Tests/test20, test21}) the results generated by the general approach $2toN$ (\classMark{TEventMaker2toN} (2toN)) and method of Lebiedowicz and Szczurek \cite{Leb-Szcz-1}, restricted to 4 particles in the final state, implemented in GenEx as \classMark{TEventMaker2to4} (2to4) are compared. To test the reaction $2p\rightarrow 2p + \pi^{+}+\pi^{-}$ at $tecm=200~\GeV$ with acceptance cuts corresponding to third row of Tab. \ref{Table-1} (\dirFileMark{test3}) were used. The results $\sigma(2to4) = 1.523  \pm 0.003 \mikrob $ and $\sigma(2toN) = 1.525  \pm 0.004 \mikrob$ agree perfectly within the statistical errors provided by Foam. 
\subsubsection{Comparison between different strategies of phase space generation for 5-body processes}\label{exploration}
Comparison of the generator strategies \classMark{TEventMakerCylindrical}(CPS) and \classMark{TEventMaker2toN} (2toN) for generation of 5-body phase space for process $pp\rightarrow pp\pi\pi\pi$ using the matrix element with Gaussian distribution of the transverse momenta of all particles (\configOptionMark{weightStrategy=2 }) and rapidity $y$ of centrally produced particles $\vert y \vert \leq 2.0$ was performed for $tecm = 200.0~\rm{GeV}$ (\dirFileMark{/Tests/test15, test16}). The phase space integrals $\sigma(2toN) = 6.70\cdot 10^{-18} \pm  7\cdot 10^{-21} \milib$ and $\sigma(CPS) = 6.67\cdot 10^{-18} \pm  8\cdot 10^{-21} \milib$ agree very well. However, if an additional cut  $ 0.01 ~\GeV^{2} \leq |t_{1},t_{2}| \leq 0.06 ~\GeV^{2}$ is imposed (\dirFileMark{/Tests/test17, test18}) the results from  CPS and 2toN strategies diverge: $\sigma(2toN) = 2.18\cdot 10^{-19} \pm  2\cdot 10^{-22} \milib$ and $\sigma(CPS) = 4.22\cdot 10^{-20} \pm 6\cdot 10^{-23} \milib$. In this case an agreement between 
CPS and 2toN strategies can be achieved if the acceptance 
cuts in the exploration phase are left as in \dirFileMark{test15} and only in the generation phase additional acceptance cut on the momentum transfer is imposed. Then phase space integral is (\dirFileMark{test19}) $\sigma(CPS) = 2.11\cdot 10^{-19} \pm  2\cdot 10^{-22} \milib$, close to $\sigma(2toN)$ result. 
\par
This example demonstrates another case of a problem with very restrictive acceptance cuts discussed earlier in subsection \ref{stability}.
In order to understand how restrictive the additional cut is one should notice that the ratio of events that  fall into the volume described by generation cuts to those generated in the volume restricted by exploration cuts is equal $0.03215$. 

\section{Summary}
A simple structure for generating events and Monte Carlo integration in particle physics was presented. The generator is specially designed for simulation of central exclusive particle production in peripheral processes e.g. double pomeron exchange. It employs the self-adaptive Monte Carlo algorithm Foam implemented in ROOT - a very powerful tool which makes it possible to generate effectively processes described by sharply peaked matrix elements and to impose restrictive cuts on the phase space. The program is characterized by a modular structure and  can be easily extended by adding new strategies for generating phase space events and calculating the matrix elements. The generator can be treated as a scheme for creating small, effective generators for various purposes. 

\appendix
\section{Phase space calculation formulae}
\label{A}
For $n=3$ particles in final state the expression (\ref{n-body}) can be transformed into (\ref{3-body-eq}) in the 
following steps:
\begin{itemize} 
\item rewrite (\ref{n-body}) in terms of the transverse momentum $\mathbf{p_{3t}}$ and pseudorapidity $y$ of particles $3$ using relation $\mathrm{d}p_z=Edy$; 
\item
introduce cylindrical coordinates $\mathbf{p}=(|p_{t}|,\phi,p_{z})$ for $\mathbf{p_{1}}$ and $\mathbf{p_{2}}$;
\item
eliminate $\mathrm{d}p_{z2}$ and $\mathrm{d}^2\mathbf{p_{3t}}$ using constraints imposed by Dirac $\delta^4(...)$;
\item 
Eliminate $\mathrm{d} p_{1z}$ using formula 
\begin{equation}\label{delta}
\delta(f(x))=\sum_{i=1}^{\#\{x_{i}: f(x_{i})=0\}} \frac{\delta(x)}{\left|\frac{df(x)}{dx}\right|}|_{x=x_{i}},
\end{equation} 
where $\#$ is a function that gives the number of elements in the set. The summation runs over the roots of the energy conservation equation, subject to additional constraint from longitudinal momentum conservation
\begin{equation}
\left\{
 \begin{array}{l}
  f(p_{1z}):=\sqrt{s}-\sqrt{p_{1z}^{2}+m_{1t}^{2}}-\sqrt{p_{2z}^{2}(p_{1z})+m_{2t}^{2}}-E_{3}=0 \\
  p_{2z}(p_{1z})=p_{az}+p_{bz}-p_{1z}-p_{3z},
 \end{array}
 \right.
 \label{energy-conservation}
\end{equation}
where $\sqrt{s}=E_{a}+E_{b}$ and  $m_{ti}=\sqrt{m_{i}^{2}+p_{ti}^{2}}$ for $i\in\{1,2\}$. 
\end{itemize}
The above system (\ref{energy-conservation}) effectively simplifies to the second order algebraic equation for $p_{1z}$ which has two solutions. One of them corresponds to backscattering of particles $1$ and $2$. As the generator is designed for simulation of the peripheral processes, it does not probe points in this branch of the phase space, because there the matrix element is negligible.
\par
Taking into account formulae (\ref{delta}) and 
\begin{equation}
 \mathcal{J}=\frac{\partial f}{\partial p_{1z}}=\frac{p_{2z}(p_{1z})}{E_{2}}-\frac{p_{1z}}{E_{1}}
\end{equation}
 one arrives at Eq. (\ref{3-body-eq}).
\par
Three-body phase space with four-momentum conservation is described by 5 independent variables, thus Foam  initialized with \configOptionMark{kDim=5} provides a vector $\mathcal{R}$ of $5$ random variables $\mathcal{R}_i, i=1,...5$ uniformly distributed on the interval $[0.0,1.0]$, which are linearly transformed to generator variables, e.g., $y=y_{\min}+(y_{\max}-y_{\min})\mathcal{R}_{1}$. The weight returned to Foam can be factorized in the following way  $\mathcal{W} = \mathcal{W}_{PhS} \cdot \mathcal{W}_{ME}$. 
The factor $\mathcal{W}_{PhS}$ due to the phase space and the affine transformation 
$\mathcal{T}: \mathcal{R}\rightarrow \{|\mathbf{p_1}|,|\mathbf{p_2}|, y,\phi_1, \phi_2 \}$ is 
\begin{equation}\label{3-body-wgt1}
 \mathcal{W}_{PhS}=|\mathcal{T}||\mathcal{J}|^{-1}(p_{1t}p_{2t},\phi_1,\phi_2,y)_{|root}\frac{1}{(2\pi)^{5}}\frac{1}{2E_{1}}\frac{1}{2E_{2}}\frac{1}{2}p_{1t}p_{2t},
\end{equation}
where
\begin{equation}
|\mathcal{T}|=|p_{t1\max}-p_{t1\min}||p_{t2\max}-p_{t2\min}||y_{\max}-y_{\min}|(2\pi)^2.
\end{equation}
The second factor $\mathcal{W}_{ME}$ due to the square of matrix element and the flux is
\begin{equation}
\mathcal{W}_{ME}=\frac{|M|^{2}}{2s}.
\end{equation}
Similarly, in the case of four-body phase space described in \ref{4-body} after analogical 
calculations one arrives at 
\begin{equation}\label{4-body-wgt1}
  \mathcal{W}_{PhS}=|\mathcal{T}||\mathcal{J}|^{-1}(p_{1t}p_{2t},\phi_1,\phi_2,y)_{|root}\frac{1}{(2\pi)^{8}}\frac{1}{2E_{1}}\frac{1}{2E_{2}}\frac{1}{2^{4}}p_{1t}p_{2t},
\end{equation}
\begin{multline}
 |\mathcal{T}|=|p_{t1\max}-p_{t1\min}||p_{t2\max}-p_{t2\min}||p_{mt\max}-p_{mt\min}|\\
|y_{3\max}-y_{3\min}||y_{4\max}-y_{4\min}|(2\pi)^3,
\end{multline}
where $\{|\mathbf{p_{1t}}|,|\mathbf{p_{2t}}|,\phi_1, \phi_2 \, y_3, y_4, |\mathbf{p_{mt}}| :=|\mathbf{p_{3t}} - \mathbf{p_{4t}}|, \phi_m \}$ variables were used. 
Finally, in case of N-body phase space described in \ref{N-body} respective expressions are (compare to \ref{3-body-wgt1})
\begin{equation}\label{N-body-wgt1}
  \mathcal{W}_{PhS}=|\mathcal{T}||\mathcal{J}|^{-1}(p_{1t}p_{2t},\phi_1,\phi_2,y)_{|root}\frac{1}{(2\pi)^{2}}\frac{1}{2E_{1}}\frac{1}{2E_{2}}\frac{1}{2}p_{1t}p_{2t}\cdot \mathcal{W}_{decay},
\end{equation}
\begin{equation}
 |\mathcal{T}|=|p_{t1\max}-p_{t1\min}||p_{t2\max}-p_{t2\min}||M^2_{\max}-M^2_{\min}||y_{\max}-y_{\min}|(2\pi)^2,
\end{equation}
where $\mathcal{W}_{decay}$ is the phase space weight for decay of object of mass $M$ into $n=N-2$ central particles calculated 
according to the recurrence relation \cite{Pilkuhn} (see also \cite{JamesCERN})
\begin{multline}
\dif^nLIPS(M^2;p_1,p_2,...p_{n-1},p_n)= \frac{1}{2\pi}\dif^2LIPS(M^2;p_d,p_n) \\\times \dif^{n-1}LIPS(s_d;p_1,p_2...p_{n-1})ds_d,
\end{multline}
where $p_d=p_1+p_2+...p_{n-1}$ and $s_d=p_d \cdot p_d$.
As already mentioned, particles $1$ and $2$ are assumed to be peripherally scattered, i.e., they have relatively small transverse momenta.

\section*{Acknowledgments}
This work was supported in part by Polish National Science Centre grants:
UMO-2011/01/M/ST2/04126 and UMO-2012/05/B/ST2/02480.




\end{document}